\documentclass[conference,10pt]{IEEEtran}
\IEEEoverridecommandlockouts
\usepackage{cite}
\usepackage{xcolor,colortbl}

\def\BibTeX{{\rm B\kern-.05em{\sc i\kern-.025em b}\kern-.08em
    T\kern-.1667em\lower.7ex\hbox{E}\kern-.125emX}}

\usepackage[utf8]{inputenc}
\usepackage[T1]{fontenc}

\usepackage{amsmath,amssymb,amsfonts}
\usepackage[abbreviations]{foreign}
\usepackage{algorithm}
\usepackage{algpseudocode}
\usepackage{graphicx}
\usepackage[margin=3pt,skip=3pt,belowskip=0pt,font={small,stretch=0.9},labelfont=bf]{caption}
\usepackage[compact]{titlesec}
\usepackage{microtype}
\usepackage{setspace}
\usepackage{listings}
\usepackage{multirow}
\usepackage{scrextend}
\usepackage{subcaption}
\usepackage{textcomp}
\usepackage{booktabs}
\usepackage{xspace}
\usepackage{url}
\usepackage{pifont}% http://ctan.org/pkg/pifont
\usepackage[colorlinks=true,linkcolor=blue,anchorcolor=blue,citecolor=blue,filecolor=black,menucolor=black,runcolor=black,urlcolor=blue]{hyperref}
\usepackage{lipsum}

\usepackage[inline]{enumitem}
\setlist{noitemsep,topsep=0pt,parsep=0pt,partopsep=0pt}

\usepackage[capitalise]{cleveref}

\usepackage{tikz}

\usepackage{fontawesome5}
\newboolean{showcomments}
\setboolean{showcomments}{true}
\ifthenelse{\boolean{showcomments}}
{ \newcommand{\mynote}[3]{
		\fbox{\bfseries\sffamily\scriptsize#1}
		{\small$\blacktriangleright$\textsf{\emph{\color{#3}{#2}}}$\blacktriangleleft$}}}
{ \newcommand{\mynote}[3]{}}

\definecolor{darkgreen}{rgb}{0.3,0.5,0.3}
\definecolor{darkblue}{rgb}{0.3,0.3,0.5}
\definecolor{darkred}{rgb}{0.5,0.3,0.3}
\newcommand{\pmem}{\textsc{PM}\xspace}

\newcommand{\sys}{\textsc{Plinius}\xspace}

\newcommand{\sgxrom}{\textsc{sgx-romulus}\xspace}
\newcommand{\sgxdnet}{\textsc{sgx-darknet}\xspace}
\newcommand{\ml}{\textsc{ML}\xspace}

\newcommand{\copyrighttext}{  \scriptsize \textcopyright 2021 IEEE.               
	Personal use of this material is permitted.                                 
	Permission from IEEE must be obtained for all other uses,                   
	in any current or future media, including reprinting/republishing this      
	material for advertising or promotional purposes, creating new collective   
	works, for resale or redistribution to servers or                           
	lists, or reuse of any copyrighted component of this work in other works.   
	Pre-print version. Published in the 51st Annual IEEE/IFIP International Conference on Dependable Systems and Networks (DSN). %For the final version, refer to DOI \href{https://doi.org/10.1109/TDSC.2020.3024562}{10.1109/TDSC.2020.3024562}}
}
\begin{document}

%\title{Plinius: Shield, Train and Persist\\ Machine Learning Models with Intel SGX and PM}
\title{Plinius: Secure and Persistent Machine Learning Model Training}
%\author{Research Paper}

\author{\IEEEauthorblockN{Peterson Yuhala}
\IEEEauthorblockA{\textit{University of Neuchâtel} \\
Neuchâtel, Switzerland \\
peterson.yuhala@unine.ch}
\and
\IEEEauthorblockN{Pascal Felber}
\IEEEauthorblockA{\textit{University of Neuchâtel} \\
Neuchâtel, Switzerland \\
pascal.felber@unine.ch}
\and
\IEEEauthorblockN{Valerio Schiavoni}
\IEEEauthorblockA{\textit{University of Neuchâtel} \\
Neuchâtel, Switzerland \\
valerio.schiavoni@unine.ch}
\and
\IEEEauthorblockN{Alain Tchana}
\IEEEauthorblockA{\textit{ENS Lyon, France} \\
Inria
}
alain.tchana@ens-lyon.fr
}
%\vspace{-30pt}

\newcommand{\copyrightnotice}{\begin{tikzpicture}[remember picture,overlay]       
	\node[anchor=south,yshift=2pt,fill=yellow!20] at (current page.south) {\fbox{\parbox{\dimexpr\textwidth-\fboxsep-\fboxrule\relax}{\copyrighttext}}};
	\end{tikzpicture}
}
\maketitle
\copyrightnotice

\thispagestyle{plain} % page numbering
\pagestyle{plain}     % page numbering

%!TEX root = main.tex

\begin{abstract}	
%\lipsum[1-2]
With the increasing popularity of cloud based machine learning (\ml) techniques there comes a need for privacy and integrity guarantees for \ml data.
In addition, the significant scalability challenges faced by DRAM coupled with the high access-times of secondary storage represent a huge performance bottleneck for \ml systems.
While solutions exist to tackle the security aspect, performance remains an issue.
%Many systems have been proposed that solve the security problem but no system so far integrates efficient solutions to both problems presented. 
Persistent memory (\pmem) is resilient to power loss (unlike DRAM), provides fast and fine-granular access to memory (unlike disk storage) and has latency and bandwidth close to DRAM (in the order of ns and GB/s, respectively).
We present \sys, a \ml framework using Intel SGX enclaves for secure training of \ml models and \pmem for fault tolerance guarantees.
\sys uses a novel mirroring mechanism to create and maintain \emph{(i)} encrypted mirror copies of \ml models on \pmem, and \emph{(ii)} encrypted training data in byte-addressable \pmem, for near-instantaneous data recovery after a system failure.
Compared to disk-based checkpointing systems, \sys is 3.2$\times$ and 3.7$\times$ faster respectively for saving and restoring models on real \pmem hardware, achieving robust and secure \ml model training in SGX enclaves.
\end{abstract}
%with efficient fault tolerance guarantees.  

%\begin{IEEEkeywords}
%TODO
%\end{IEEEkeywords}

%!TEX root = main.tex
\section{Introduction}

\begin{figure}[!t]
	\centering
	\includegraphics[scale=0.8]{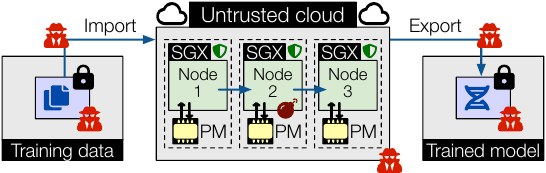}
	\caption{\ml and model building over untrusted public clouds. In case of failures or job pre-emption ({\small\faBomb}), the current state should be securely persisted to \pmem. Without proper mitigations, data and trained models could leak ({\small\faUserSecret}).}
	\label{fig:untrustedml}
\end{figure}

Privacy-preserving machine-learning~\cite{ppmlSP19} is a challenging computational paradigm and workflow.
Data and computation must be protected from several threats, \eg, powerful attackers, compromised hypervisors and operating systems, and even malicious cloud or human operators~\cite{stealing,slalom}.
Preserving the confidentiality of the models (\ie, weights and biases) being trained, as well as the input datasets is paramount: these are the most valuable business assets. 
Example application domains include health, finance, Industry 4.0, \etc.
Given the significant amount of computing resources typically required during the training phase, moving the model training over public clouds appears to be a pragmatic approach.
However, this immediately leads to contradictory arguments.
On the one hand, one benefits from the endless scalability and dependability features of public clouds, as well as pushing away valuable assets from potentially compromised on-premises infrastructures to the cloud. On the other hand, exposing confidential datasets and models to untrusted clouds must be avoided.
\cref{fig:untrustedml} shows our target scenario, \ie, training of ML models over untrusted public clouds, showcasing the threats that this scenario implies.

Trusted execution environments (TEE) are quickly becoming the go-to solution to tackle such confidentiality requirements, and several cloud providers nowadays offer TEE-enabled computing instances (IBM,\footnote{\url{https://www.ibm.com/cloud/data-shield}} Azure\footnote{\url{https://azure.microsoft.com/en-us/solutions/confidential-compute/}}).
Intel software guard extensions (SGX)~\cite{vcostan} is a TEE that offers applications secure memory regions called \emph{enclaves} to shield code and data from unwanted accesses. 
SGX is a promising candidate for protecting applications, given its wide availability across a variety of cloud providers.
However, SGX imposes security restrictions on enclave code (\eg, disallowing system calls), typically requiring application-level changes, as well as limited memory capacity of SGX enclaves, which requires developers to minimize the \emph{trusted computing base} (TCB). 

%then, we say that ML jobs 
While offloading ML training jobs to SGX-enabled clouds might solve the confidentiality issue, such jobs are typically deployed in batch.  
Typically, batch jobs have lower priorities than latency sensitive (\eg, production, user-facing) services~\cite{iorgulescu2018perfiso}.
In order to avoid resource waste because of workload variations, ML applications are colocated with latency sensitive applications.
The former are automatically killed when the latter needs more resources.
Another practice which may lead to interruptions of ML jobs is the use of cheap yet unreliable virtual machine instances such as EC2 spot instances~\cite{wang2018empirical}.
The latter are automatically terminated when a better offer (\ie, spot price) is made by another user.
To avoid restarting model training from scratch when a task is killed, one can checkpoint/restore the current model on persistent storage.
For instance, AWS SageMaker~\cite{liberty2020elastic} suggests to frequently checkpoint the models to avoid abrupt data-loss when using spot instances.
\cref{fig:mlpipeline} represents this general \ml pipeline.
%\footnote{\url{https://docs.aws.amazon.com/sagemaker/latest/dg/model-managed-spot-training.html}}
%\vs{add a drawing here to show this: The overall workflow is given in figure xxx.}
However, frequent checkpointing on secondary storage leads to significant I/O overheads, and relying on volatile memory (\ie, DRAM) to mitigate these overheads would prevent to resume the job in case of task eviction.
%We also observe how ML training is typically operated over cheap yet unreliable instances (\eg, EC2 Spot instances~\cite{yi2010reducing}). 
%For instance, AWS SageMaker\footnote{\url{https://docs.aws.amazon.com/sagemaker/latest/dg/model-managed-spot-training.html}} suggests to frequently checkpoint the models to circumvent abrupt data-loss, leading however to the mentioned drawbacks.
%\vs{here we need to mention the EC2 Managed Spot Training scenario with AWS SageMaker, 
%\vs{here we would need a small plot showing this.}\py{we could refer the reader to Fig \ref{fig:crash}(b). Thats what happens when you cannot resume where you left off..}

%now we introduce M
Emerging memory technologies like \emph{persistent memory} (\pmem)~\cite{yang2020empirical} have the potential to address the significant scalability challenges faced by DRAM, as well as the high latency of secondary storage. 
\pmem is persistent on power failure, byte-addressable, and can be accessed via processor \texttt{load} and \texttt{store} instructions. 
Recent work~\cite{yang2020empirical} shows how on-the-market \pmem solutions such as Intel Optane DC \pmem~\cite{yang2020empirical} result in significant performance gains for various applications. Cloud services like MS Azure already provide \pmem offerings~\cite{azurepm}, and we expect this technology to gain even more momentum. 
However, using \pmem in privacy-preserving ML jobs opens additional security risks: confidential model parameters could be persisted in plain-text on \pmem, or possibly be exposed at runtime to malicious privileged users or compromised operating systems. 
We take the stance that there is the need to develop tools and mechanisms to enable these applications to leverage \pmem in such secure computation environments. 
%A very promising field to benefit from the potential of \pmem is machine learning (\ml).
%In spite of the appealing properties of \pmem, using the latter in applications introduces security challenges associated with data remanence~\cite{zuo2018secpm}, because sensitive data could be persisted in plaintext prior to a (possibly malicious) system crash.
%At the same time, many of the applications to benefit from \pmem require completely shielded environments to execute. 
%For instance, in the context of privacy preserving \ml applications, models built on public cloud infrastructures are susceptible to several types of security and privacy issues~\cite{stealing,modelinv,membership,oblivious,mlcapsule,chiron,slalom}. 
%In this work we focus specifically on preserving the confidentiality of model parameters (\ie weights and biases) and training or inference data sets, because these typically represent the most valuable assets.
In this work we build the first framework that integrates secure \ml with Intel SGX with fault tolerance on \pmem.
%deal with confidential models liable to risk of intellectual property leaks, or perform data analytics on sensitive data~\cite{mohassel2017secureml,papernot2018sok}
%\vs{here we need a sentence saying that pure software-based approaches are not sufficient/too slow/unpractical. THen, we introduce TEEs}
%\vs{to redo. Figure showing the general scenario that we tackle: training of ML models over untrusted public clouds. Figure~\ref{fig:untrustedml}}
State-of-the-art \pmem libraries (\eg, Intel Persistent Memory Development Kit~\textsuperscript{\ref{pmdk-link}} , Romulus~\cite{romulus}, \etc.), as well as \ml frameworks (\eg, Tensorflow~\cite{abadi2016tensorflow}, Darknet~\cite{darknet}, \etc), require considerable porting efforts to be fully functional within SGX enclaves.
%For instance, Intel PMDK uses \vs{XXX} of such forbidden operations.
Tools exist (\eg, library OSes like Graphene-SGX~\cite{graphene} and Occlum~\cite{shen2020occlum}, or modified enclave-compatible C libraries like SCONE~\cite{scone}) to run unmodified applications inside SGX enclaves, at the downside of larger TCB sizes (thus larger attack surfaces) and large memory footprint, thus reducing performance.
In \ml scenarios with large confidential models and data sets, enclave memory becomes a major bottleneck that only important engineering efforts could optimize. % be optimized in such scenarios. 
%There is no need to bloat enclave memory with a full library OS for example; having just the minimal necessary code in the TCB is sufficient.
%While manually porting applications into Intel SGX is often a challenging endeavour, it typically achieve improve performance, especially dealing with memory-constrained applications\vs{we need refs to support this}.

We present \sys, a secure \ml framework that leverages \pmem for fast checkpoint/restore of machine learning models.
%especially optimized for privacy-preserving \ml. 
\sys leverages Intel SGX to ensure confidentiality and integrity of \ml models and data during training, and \pmem for fault tolerance. 
\sys employs a \emph{mirroring mechanism} which entails creating an encrypted mirror copy of an enclave model directly in \pmem. The mirror copy in \pmem is synchronized with the enclave model across training iterations.
\sys maintains training data in byte addressable \pmem. 
Upon a system crash or power failure while training, the encrypted \ml model replica in \pmem is securely decrypted in the enclave, and used as next starting point of the training iteration: the training resumes where it left off, using training data already in memory. This avoids costly serialization operations of disk-based solutions.
To validate our approach, we build and contribute \sgxdnet, a  complete port of Darknet \ml framework~\cite{darknet} to SGX, as well as \sgxrom, an SGX-compatible \pmem library on top of an efficient PM library~\cite{romulus}.
We compare \sgxrom with unmodified Romulus library running in a SCONE container and our results show \sgxrom is best suited for \sys framework.
Using \sys, we build and train \emph{convolutional neural network} (CNN) models with real world datasets (\ie, MNIST~\cite{mnist}) and show \sys reduces overhead by $\sim$3.5$\times$ for model saving, and $\sim$2.5$\times$ for model restores with real SGX hardware and emulated \pmem.\footnote{Machines with SGX and \pmem support not on the market yet (Oct/2020).}

In summary, we make the following contributions: 
\begin{itemize}[leftmargin=*]
\item We implement and release as open-source \sgxrom\footnote{\url{https://github.com/Yuhala/sgx-romulus}} on top of Romulus~\cite{romulus} for Intel SGX. \sgxrom manipulates \pmem directly from within SGX enclaves, without costly enclave transitions between secure and unsecured parts of an SGX application. 

\item \sloppy{We design, build, and release as open-source, \sgxdnet\footnote{\url{https://github.com/Yuhala/sgx-dnet}} an extension of Darknet~\cite{darknet} for Intel SGX.
\sgxdnet can perform secure training and inference on \ml models directly inside SGX enclaves.} %We make \sgxdnet available at~\cite{sgx-dnet}.
\item We present \sys, an open-source framework\footnote{\url{https://github.com/Yuhala/plinius}} that leverages \sgxrom and \sgxdnet to provide an end-to-end fault tolerance mechanism to train models in privacy preserving \ml settings. %It is available as open-source at \cite{plinius}. %\sys exploits Intel SGX for confidentiality and integrity, based on a novel mirroring mechanism, detailed in \S\vs{FIX}, to maintain an encrypted twin copy of a model in \pmem, and periodically synchronizing copies during training iterations.
\item We provide a comprehensive evaluation of \sys, using real PM hardware and real AWS Spot traces, showing its better performance when compared with traditional checkpointing on secondary storage (\ie disk or SSD)
\end{itemize}

%\at{(1) Rappeler que les expés sont sur real PM. (2) qu'on a construit un real scenario avec des real traces AWS.}
\smallskip\noindent\textbf{Roadmap.} This paper is organized as follows.
\S\ref{sec:background} describes background concepts, while \S\ref{sec:tmodel} presents our threat model.
\S\ref{sec:architecture} presents the architectures of \sgxrom, \sgxdnet and \sys, while \S\ref{sec:implementation} digs into the implementation details. 
The experimental evaluation of our system is in \S\ref{sec:evaluation}. 
We discuss related work in \S\ref{sec:related}, before concluding and hinting at future work in \S\ref{sec:conclusion}.
%and a detailed analysis of the results

% \input{motivations}
%!TEX root = main.tex
\section{Background}
\label{sec:background}

This section presents a background on Intel SGX,  \pmem, as well as some machine-learning concepts specific to \sys.

\smallskip\noindent\textbf{Intel software guard extensions (SGX)}~\cite{vcostan} is a set of extensions to Intel's architecture that permits applications to create CPU-protected memory areas (\ie, \emph{enclaves}) shielding confidential code and data from disclosure and modifications. 

SGX reserves a secure memory region called the \emph{enclave page cache} (EPC) for enclave code and data. The processor ensures that software outside the enclave (\eg, the OS kernel or hypervisor) cannot access EPC memory. The enclave can access both EPC and non-EPC memory. 

Data in the EPC is in plaintext only in on-chip caches and is encrypted and integrity-protected in the \emph{memory encryption engine} (MME) once it is evicted from the cache to memory.  
Current Intel processors support a maximum of 128\,MB of EPC memory, of which 93.5\,MB is usable by SGX enclaves. This limits the total size of code and data allowed within the EPC. % be loaded into the enclave during execution. 
To support applications with larger memory needs, the Linux kernel provides a paging mechanism for swapping pages between the EPC and untrusted memory.

Enclaves cannot issue system calls and standard OS abstractions (\eg, file systems, network), which are ubiquitous in real world applications. 
All system services thus require costly enclave transitions, up to 13'100 CPU cycles~\cite{sgxperf}. 
The Intel SGX application design requires splitting applications into a trusted (the enclave) and untrusted part. 
To achieve communication across the enclave boundary, the Intel SGX SDK provides specialized function call mechanisms,\ie, \emph{ecalls} and \emph{ocalls}, respectively to enter and exit an enclave~\cite{sgxdevref}.
 
%Trusted code inside the enclave uses \emph{ocalls} to access routines in the untrusted portion of an application, and code in the untrusted portion leverages \emph{ecalls} to access trusted enclave functionality. 
To mitigate security risks, the TCB should be as small as possible.
Systems exist to run unmodified applications inside enclaves, either by porting entire library OSes into the enclave~\cite{graphene,shen2020occlum} or via a modified \texttt{libC} library, specialized for containerized services~\cite{scone}.
%Haven and Graphene-SGX port entire library OSes into the enclave, leading to large TCBs of millions of LOC. 
%SCONE uses a modified C standard library to support containerized services inside enclaves. 
These solutions are efficient with small application binaries, but quickly show limits for memory-constrained applications such as \ml.

\smallskip\noindent\textbf{Persistent memory and PM libraries.}
\pmem is a  novel  memory technology that is \emph{non-volatile}, \emph{byte-addressable}, and has latency and bandwidth similar to that of DRAM. 
\pmem resides on the memory bus and can be accessed directly using CPU \texttt{load} and \texttt{store} instructions. 
Intel Optane DC \pmem is commercially available since April 2019. 
Optane DC \pmem scales better than DRAM and hence provides much larger capacity (up to 512\,GB per PM module).
Intel Optane DC \pmem modules can operate in two modes: \emph{memory mode} where they are simply used to expand main memory capacity without persistence, and \emph{app direct mode} which provides persistence~\cite{yang2020empirical}. \sys leverages \pmem in \emph{app direct mode}.

Applications can leverage \pmem in \emph{app direct mode} by using standard operating system calls (\ie \textit{read,write}) through the file system in the same way slower storage devices like SSDs and HDDs are accessed. This improves application performance but does not leverage the load/store interface provided by the \pmem modules. In \sys, we enhance the \ml library to do direct loads/stores from/to \pmem. This configuration is more challenging as it requires application modification. However, it results in more significant performance gains since persistent updates bypass both the kernel and file system~\cite{izraelevitz2019basic}.

Server-grade CPUs natively support up to 3\,TB of \pmem~\cite{pmemindexes}, hence revealing \pmem as an attractive solution for fault tolerant applications.
%Notwithstanding its very appealing features, 
%\pmem has a higher read/write latency than DRAM~\cite{nvsl,pmemindexes} making DRAM a better option in terms of performance. 
%Current non-volatile architectures~\cite{sytare} combine DRAM and \pmem to exploit the benfits from both technologies.
Due to data remanence~\cite{zuo2018secpm}, using \pmem could introduce security risks, in particular for confidentiality and data integrity.

\begin{figure}[t!]
	\centering
	\includegraphics[scale=0.68,trim={0 0 0 0}]{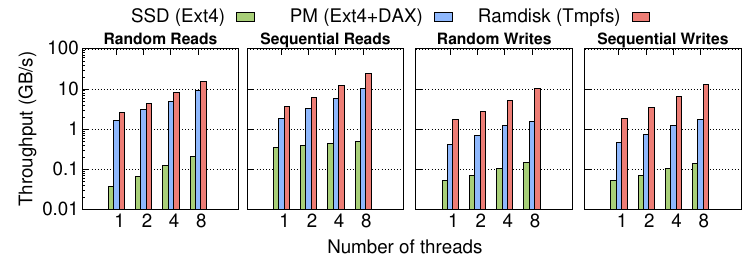}
	\caption{Read/write throughput for sequential/random workloads on SSD, PM and Ramdisk using the \texttt{sync} I/O engine on FIO. 512\,MB file per thread, 4\,KB block size. Write workloads issue an \texttt{fsync} for each written block, average over 3 runs.}
	\label{fig:pmem-perf}
\end{figure}

The use of \pmem requires a paradigm shift for application developers. 
Several software tools and libraries have been proposed, such as Romulus~\cite{romulus}, Mnemosyne~\cite{volos2011mnemosyne} or Intel's PMDK\footnote{\label{pmdk-link}\url{https://pmem.io/}} to facilitate \pmem related development. 
%Using \pmem libraries for \pmem related development is advantageous in that the latter provide a simple and familiar programming interface for programmers to manage persistent data.
\pmem libraries expose \pmem to applications by memory-mapping files on a persistent memory-aware file system  with \emph{direct access} (DAX) capabilities. 
DAX removes the OS page cache from the I/O path and allows for direct access to \pmem with byte-granularity.\textsuperscript{\ref{pmdk-link}} 

To characterize our PM units, we execute FIO\footnote{\url{http://freshmeat.sourceforge.net/projects/fio}} with sequential and random workloads, and compare the read and write throughputs for native \texttt{Ext4} over an SSD drive, \texttt{Ext4+DAX} on PM, and a \texttt{tmpfs} partition over volatile DRAM.
%\cref{fig:pmem-perf} reports our results.
We observe (\cref{fig:pmem-perf}) that the DAX-enabled file system on PM performs consistently better than its non-DAX counterpart on SSD, and is close to RAM-tmpfs performance (in the order of GB/s).

Specific processor instructions (\ie, \texttt{CLFLUSH}, \texttt{CLFLUSHOPT}, \texttt{CLWB}) are used to flush data from cache lines to the \pmem memory controller. Through asynchronous DRAM refresh~\cite{yang2020empirical} data in the memory controller's write buffers is guaranteed to be persisted in \pmem in case of a power failure. 
Persistence fences (\ie, \texttt{SFENCE}) guarantee consistency by preventing \texttt{store} instructions from being re-ordered by the CPU.
%Persistence fences are slow and represent a bottleneck in \pmem libraries.~\cite{romulus}. 
\pmem libraries like Romulus and the PMDK provide transactional API which enable developers to perform atomic updates on persistent data structures. 

Romulus provides durable transactions via twin copies of data in \pmem and relies on a volatile log to track memory locations being modified in a transaction.
The first copy, called the \texttt{main} region, is where user-code executes all in-place modifications; the second copy, the \texttt{back} region, is a \emph{backup} (or snapshot) of the previous consistent state of the \texttt{main} region. 
Following a  crash while mutating \texttt{main}, the content of \texttt{back} is restored to \texttt{main}. % thus restoring the latter to its previous consistent state. 
%Romulus uses a volatile log to track memory locations (on \texttt{main}) being modified in a transaction. Following a successful atomic update on \texttt{main}, the contents of the modified memory locations are replicated on \texttt{back}.\vs{what happens when a failure happens while replicating on back?} 
Romulus uses at most four persistence fences for atomic updates on data structures, regardless of transaction size, and a \emph{store interposition} technique to ensure cache lines are correctly flushed to \pmem.
The design of Romulus permits to have low \emph{write amplification}~\cite{romulus} relative to other PM libraries, and hence we build on top of it as \pmem library, by porting it to be SGX-compatible.
%Romulus uses \emph{store interposition}~\cite{romulus} to ensure cache lines are correctly flushed to \pmem following a \texttt{store} operation on persistent data. 
%\vs{detailed description of redo logs could be moved to appendix if needed}
%Popular \pmem libraries like Mnemosyne and PMDK and failure atomicity mechanisms like \emph{Atlas}~\cite{chakrabarti2014atlas} and \emph{JustDo}~\cite{justdo} use persistent \emph{undo}~\cite{Coburn2012,justdo} and \emph{redo}~\cite{volos2011mnemosyne,justdo} log mechanisms for transactional persistence. The latter mechanisms lead to additional bytes written to \pmem for every byte of user data stored on \pmem during a transaction, a phenomenon referred to as \emph{write amplification}~\cite{whisper}. Persistent undo and redo logs equally increase the number of persistence fences required per transaction. The Romulus design permits the latter to have minimal persistence fences and write amplification when compared to other failure atomicity mechanisms and \pmem libraries, hence better performance. This is the basis for our choice of Romulus for \pmem related development.

\smallskip\noindent\textbf{Training ML models.}
%The main concept of machine learning is to develop algorithms that can analyse large sample data sets and build mathematical models which reflect general relationships in the sample data.
A \ml \emph{model} can be described as a function that maps an input to a target output based on a set of parameters~\cite{chiron}.
A linear regression model for example is a function $\mathbf{f(x)} = \mathbf{W}^\intercal \mathbf{x} + \mathbf{b}$ where $\mathbf{W}$ represents the model weights, $\mathbf{b}$ the bias vector, and $\mathbf{x}$ the input vector.
The weights and biases are the learnable parameters of a model, and they determine the output of the latter for a given input. 

The goal of model training is to obtain the set of learnable parameters that minimizes a \emph{loss function} and maximizes the model's accuracy on the training data. 
The loss function is a scalar function that quantifies the difference between the predicted value (for a given input data point) and the ground truth or real value~\cite{abadi2016tensorflow}. 
During training, the learning algorithm iteratively feeds the model with batches of training data, calculates the loss, and updates the model parameters in such a way as to minimize the loss.
A very popular learning algorithm used in \ml for loss minimization is \emph{stochastic gradient descent} (SGD)~\cite{mohassel2017secureml}.
%Training large \ml models, \eg  deep neural networks, require large training data sets and can take considerable time. %several hours, days, or even weeks
%Long running training jobs are likely to experience failures or pre-emption during the training process~\cite{borg19-eurosys20}. 
%\ml frameworks implement user-level \emph{checkpointing} for fault-tolerance. 

In this work, we rely on supervised learning, where a (costly) training phase builds a model out of labelled data, followed by a classification/inference phase using the model.
Examples include visual object recognition, spam filtering, \etc. 
%and this workflow is supported by state-of-the-art frameworks such as Tensorflow~\cite{abadi2016tensorflow}, Caffe~\cite{caffe}, and Darknet~\cite{darknet}.

\begin{figure}[!t]
	\centering
	\includegraphics[scale=0.78]{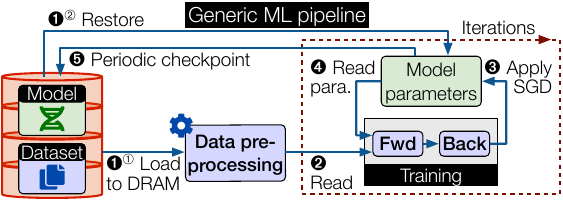}
	\caption{General machine learning pipeline: \emph{Fwd}=Forward propagation, \emph{Back}=Backward propagation.}
	\label{fig:mlpipeline}
\end{figure}

\cref{fig:mlpipeline} shows a typical \emph{supervised} \ml model training pipeline. 
Training data is read from secondary storage (\autoref{fig:mlpipeline}-\ding{202}$^{\textsuperscript{\ding{192}}}$), preprocessed and used to train the model (\autoref{fig:mlpipeline}-\ding{203},\ding{204},\ding{205}). 
Training models such as deep neural networks can take up to several days, a time window sufficiently long for training jobs to experience failures or pre-emptions~\cite{abadi2016tensorflow}.
Also, large \ml datasets (\ie, order of GBs) are very common in the training phase. 
In the event of a failure during training, the model being trained as well as the training data sets resident in DRAM are lost and need to be re-read from secondary storage upon restart. 
Several state-of-the-art \ml frameworks (\eg, Tensorflow~\cite{abadi2016tensorflow}, Darknet~\cite{darknet}, Caffe~\cite{caffe}, \etc.)  provide mechanisms to checkpoint model states to secondary storage during training. %\vs{we should have a reference here showing this is a common practice}.
However, the high latency and low bandwidth (in the order of MB/s) of secondary storage makes failure recovery a fundamental problem. 
The mentioned appealing properties of \pmem make the latter particularly interesting for fault tolerance in such \ml scenarios.

In this work, we implement \sys, a novel \ml framework which leverages \pmem for fault tolerance and Intel SGX to ensure confidentiality and integrity of \ml models, as well as sensitive training data.
We build our \ml framework on Darknet, which is popular in the \ml community, provides good performance and is easily portable to Intel SGX.
%\vs{add 1 sentence saying why we have chosen Darknet: good perfs, implemented in C/C++ and easy to port to SGX, good adoption}
%Emerging memory technologies like \pmem are with DRAM-like access times have the potential to solve such performance problems.

%\sys provides tools and mechanisms to leverage \pmem for quick failure recovery for \ml models in the context of privacy preserving \ml.   

%!TEX root = main.tex
\section{Threat Model}
\label{sec:tmodel}

\sys has three primary goals:
%The primary goals of \sys are threefold:
\emph{(1)} to ensure confidentiality and integrity of a \ml model's parameters (\eg, weights, biases) during training;
\emph{(2)} to ensure confidentiality and integrity of the model's replica on \pmem used for fault tolerance; and
\emph{(3)} to ensure confidentiality and integrity of training data in byte-addressable \pmem.

The system is designed to achieve these goals while facing a powerful adversary with physical access to the hardware and full control of the entire software stack including the OS and hypervisor. The adversary seeks sensitive information inside the enclave, on DRAM or \pmem, or data from the processor. 

Model hyper-parameters such as model architecture, number of layers, size of training batches or type of training data are usually public information, as they do not leak any information about trained model parameters or sensitive training data~\cite{grover2018privado,chiron,mlcapsule}. In order to mitigate possible threats linked to malicious data sources, \sys supports secure provisioning of model hyper-parameters via the SGX remote attestation mechanism.
%\smallskip\noindent\textbf{Assumptions.}
%We assume a that the adversary knows hyper-parameters of the model such as model architecture, number of layers, size of training batches or type of training data (\eg, images, text) because these do not leak any information about trained model parameters or sensitive training data.
We assume that the adversary cannot physically open and manipulate the processor package, that enclave code is correct and it does not leak sensitive information (\eg, encryption keys) intentionally.
%\smallskip\noindent\textbf{Further threats.}
Denial-of-service and side-channel attacks~\cite{van2018foreshadow,schwarz2017malware}, for which solutions exist~\cite{oleksenko2018varys,gruss2017strong}, are considered out of scope.

%!TEX root = main.tex

\begin{figure}[!t]
	\centering
	\includegraphics[scale=0.78]{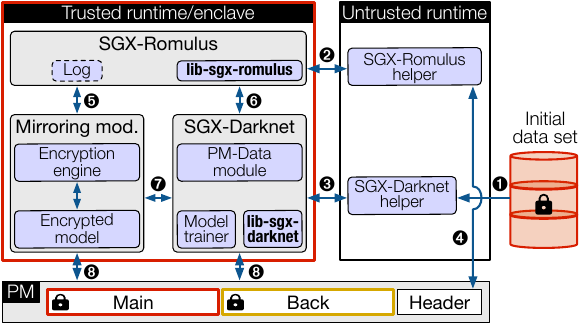}
	\caption{\sys architecture.}
	\label{fig:arch}
\end{figure}

\section{\sys Architecture}
\label{sec:architecture}

%This section presents the architecture of \sys and its general workflow, and discusses on how new \ml libraries can be integrated into \sys.

%\subsection{Architecture}
%\py{we could use this subsection to discuss why we manually port our applications instead of using a libOS like scone. This was proposed by one of the MW reviewers.}
%texDesign considerations.}
The design of \sys introduces an important issue: minimizing the TCB. A design approach based on a libOS like Graphene SGX or SCONE containers introduces thousands of lines of code into the enclave runtime, increasing security risks. Furthermore, with such a design, the enclave quickly reaches its memory limitation leading to a dramatic loss of performance.
In light of these problems and following the SGX guidelines~\cite{sgxdevref}, we design an architecture partitioned into trusted and untrusted parts. 

By manually porting the \pmem and \ml libraries via separation into trusted and untrusted components, \sys achieved a TCB reduction of $\sim$44\% in terms of LOC. 

The architecture of \sys consists of three main components interacting with each other:
\emph{(1)} an SGX-compatible PM library, \ie, \texttt{sgx-romulus};
\emph{(2)} an SGX-compatible deep-learning framework, \ie, \texttt{sgx-darknet}; and
\emph{(3)} a \texttt{mirroring module}, which synchronizes the \ml model inside the enclave with its encrypted mirror copy in \pmem.  
Figure~\ref{fig:arch} shows how these components interact. 
We detail each of them in the remainder of this section.

%~\\
\smallskip\noindent\textbf{SGX-Romulus} is a port of Romulus~\cite{romulus} to Intel SGX. 
\sgxrom implements durable transactions in \pmem directly within an SGX enclave. 
It consists of a secure \sloppy{user-space library, \texttt{lib-sgx-romulus},} which provides durable, concurrent transactions, and persistence primitives required to create and manage persistent data structures in \pmem. 
\sgxrom maintains a volatile log in enclave memory which logs the addresses and ranges of modified data in the current transaction.
The size of the log varies with transaction size. A helper library in the untrusted runtime, \texttt{sgx-romulus-helper}, communicates with \sgxrom and permits to invoke necessary system calls (\eg, \texttt{mmap,munmap}) which are required when leveraging \pmem via a DAX-enabled file system. %\vs{put numbers in the figure, we reference them in the text}. 

At application initialization, \texttt{sgx-romulus-helper} memory-maps the file corresponding to the persistent memory region (\emph{main}, \emph{back} and \emph{header}) into application virtual address space (VAS), via a \texttt{mmap} system call. 
Then, \texttt{sgx-romulus-helper} initializes the \emph{persistent header}~\cite{romulus}, which holds metadata to track the consistency state of the \emph{main} and \emph{back} regions, a reference to an array of persistent memory objects, and a pointer to the memory allocator's metadata (\eg, allocated and unallocated \pmem).
%\footnote{The adversary could send an incorrect address to the enclave, with the goal of corrupting data in \pmem. Note that if this is the goal of the adversary i.e DOS, they could still write malicious code to corrupt the \pmem regions from outside the enclave. The encrypted information in \pmem remains confidential nonetheless.}\py{not really needed: we have already assumed DOS attacks are not taken into consideration.}
%Once the enclave validates the address (\eg, checks that it is not an enclave address; this mitigates attacks which could compromise enclave integrity~\cite{van2019tale})
The address of the persistent header is passed to \sgxrom via an \texttt{ecall} and once the enclave validates this address, it then completes the \pmem region initialization, as further detailed in Algorithm~\ref{alg:sgxrominit}. 
The enclave can then create or update persistent data structures in \pmem. 
%\sgxrom uses the same synchronization primitives provided by Romulus to synchronize enclave threads spawned in the untrusted runtime.  
Upon graceful termination of the enclave application, the enclave runtime issues an \texttt{ocall} to unmap the \pmem region from application VAS via the \texttt{munmap} system call.

\smallskip\noindent\textbf{SGX-Darknet} is a port of Darknet~\cite{darknet} to Intel SGX. 
While many TEE-based ML libraries only provide support for inference, \sgxdnet supports both secure training and inference on \ml models in Intel SGX enclaves. 
In order to achieve a minimal TCB, we separate \sgxdnet into trusted and untrusted parts. Our separation strategy involves keeping out of the enclave, as much as possible, computations which do not require any particular security. Examples include parsing of model configuation files, and initial data loading into DRAM.

To minimize code changes for commonly used (but unsupported) routines in Darknet (\eg, \texttt{fread}, \texttt{fwrite} \etc), \sgxdnet redefines the former as wrapper functions for \texttt{ocall}s to the corresponding libC functions in the untrusted runtime. 
A support library in the untrusted runtime, \emph{sgx-darknet-helper}, provides the implementations of those \texttt{ocall}s invoking the corresponding libC routines. 
%The \emph{sgx-darknet-helper} module contains tools to read and parse Darknet configuration files (used by Darknet to define the model's architecture), as well as the initial encrypted training data sets from secondary storage into main memory.
The \emph{pm-data-module} is used by \sgxdnet to write/read encrypted data sets to/from \pmem.
Finally, \emph{lib-sgx-darknet} provides the API to train and do inference on models from within the enclave runtime. 
%\vs{put numbers in the figure to be referenced in the text}
\begin{figure}[!t]
	\centering
	% \hspace*{-0.55cm}
	\includegraphics[scale=0.8]{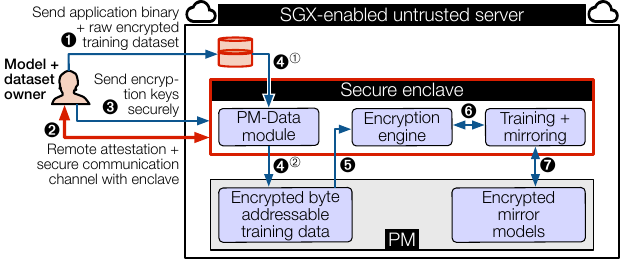}
	\caption{Full model training workflow with \sys.}
	\label{fig:workflow}
\end{figure}

\smallskip\noindent\textbf{Mirroring module.}
This component is in charge of creating and updating encrypted mirror copies of enclave \ml models on \pmem. It contains the necessary logic to instantiate models that are both persistent and directly byte-accessible via loads and stores. It leverages the transactional API provided by \sgxrom to perform atomic updates on persistent models in \pmem. This is crucial as it prevents any inconsistency in \pmem data structures in the event of a system failure during data updates.
The logic for building and managing persistent versions of complex data structures like \ml models can get very bulky, and so we preferred to build a separate module for that rather than integrate it directly into \sgxdnet. 

The \emph{encryption engine} is responsible for encrypting/decrypting model parameters to be mirrored to/from the \pmem model, as well as in-enclave decryption of encrypted training data resident in \pmem.

In-enclave symmetric encryption/decryption relies on AES Galois counter mode (GCM)~\cite{nist} implementation from the Intel SGX SDK.
%Data encryption/decryption in the enclave is done using the AES Galois counter mode (GCM)~\cite{nist} symmetric encryption algorithm provided by the Intel SGX SDK.
AES-GCM uses a 128, 192 or 256 bit key for all cryptographic operations, and provides assurance of the integrity of the confidential data~\cite{nist}.
\sys uses a 128 bit key for all cryptographic operations.
%The security of counter mode encryption algorithms as AES-GCM is based on the premise that each initialization vector (\texttt{IV}) used for encryption operations is never repeated with a given key~\cite{nist}.

As recommended by~\cite{nist}, for every encryption operation, we generate a random 12-byte \textit{initialization vector} (IV) using the \texttt{sgx\_read\_rand()}~\cite[p.~200]{sgxdevref} function from the Intel SGX SDK. The encryption algorithm divides each plain text buffer into 128 bit blocks which are encrypted via AES-GCM. 
%We therefore render our encryption operations probabilistic by generating a random 12-byte IV for every encryption operation performed.
%\vs{what is the cost to generate such random 12-bytes? this is done in-enclave?}.\py{It is done in the enclave and is inexpensive: takes 1.3 micro seconds}
The \texttt{IV} and a 16-byte message authentication code (MAC) are then appended to each encrypted data buffer. 
The MAC is used to ensure data integrity during decryption. 

The key used for encryption/decryption can be provisioned to the enclave via remote attestation~\cite[p.~99]{sgxdevref} or could be generated securely (\eg, if training data is not encrypted) inside the enclave using \texttt{sgx\_read\_rand()}. 
The encryption key, once generated or provisioned, can be securely sealed~\cite[p.~96]{sgxdevref} by the enclave for future use.

\smallskip\noindent\textbf{Full \ml workflow with \sys.} Figure \ref{fig:workflow} shows the full \ml workflow with \sys. % interpreted as follows.
The owner of the data and the model sends the application binary and raw encrypted training data to the remote untrusted server (Figure~\ref{fig:workflow}-\ding{202}).
She then performs remote attestation (RA), establishes a secure communication channel (SC) with the enclave (\autoref{fig:workflow}-\ding{203}) and sends encryption keys to the latter (Figure~\ref{fig:workflow}-\ding{204}).
The \pmem-data module transforms encrypted data on disk to encrypted byte addressable data in \pmem (Figure~\ref{fig:workflow}-\ding{205}$^{\textsuperscript{\ding{192},\ding{193}}}$).
The training module reads and decrypts (with keys obtained from RA \& SC) batches of training data from \pmem (Figure~\ref{fig:workflow}, \ding{206}-\ding{207}) with the trained model being mirrored to \pmem or into the enclave for restores (Figure~\ref{fig:workflow}-\ding{208}).

\textbf{Integration with different \ml libraries. }%in \sys
The current \sys architecture uses Darknet as the \ml library, due to its efficient and lightweight implementation in C that facilitates integration with SGX enclaves. % which is a very popular ML library. 
Other \ml libraries could be integrated into the \sys architecture. % if they are properly ported into Intel SGX.
%The ease of porting depends on the complexity of the library. 
In fact, once the \ml library is ported to SGX, the same \sys architecture holds. 
%However, the \emph{mirroring module} should be adapted to consider the newly introduced data structures, as different \ml libraries may represent models differently.

To validate the generality of our architecture, we applied our mirroring mechanism within Tensorflow~\cite{abadi2016tensorflow}, another popular \ml library.
Tensorflow uses \emph{tensor} data structures to store model information (\eg, weights and biases). 
Our implementation creates mirror copies of tensors in \pmem and restores them in enclave memory using \sys's mirroring mechanism. 
However, due to the large memory footprint of Tensorflow-based \ml applications with respect to our EPC limit (93.5\,MB), we opted to use Darknet \ml library, which is lightweight but equally efficient. 
%For instance, we measured 99\,MB for a simple Tensorflow linear regression application binary.
%This alone exceeds our EPC limit (93.5\,MB) and therefore is not suitable for an SGX-based system like \sys. 
%A similar application based on Darknet has a binary size less than 3\,MB, hence our choice of the former as \ml library.

%!TEX root = main.tex
\section{Implementation Details}
\label{sec:implementation}

We implement \sys in C and C++ and it comprises 28'450 lines of code (LOC) in total, the trusted portion being 15'900 LOC. We use Intel SGX SDK v2.8 for Linux. The total size of application binary including the enclave shared library after compilation is 3\,MB.
In the remainder, we describe further details and a rundown for a \ml model training.
%This section describes the implementation of \sys as well as a rundown of the training process for a \ml model.

\textbf{Initialization.} In this phase, \sys memory maps \pmem into application virtual address space (VAS) (see Algorithm~\ref{alg:sgxrominit}, lines 3-5) and initializes the persistent regions \emph{main} and \emph{back} (Algorithm~\ref{alg:sgxrominit}, line 12), so that both regions are consistent before the training starts. 
%Algorithm~\ref{alg:sgxrominit} summarizes \sys's initialization steps (\texttt{init\_sgx\_romulus} and \texttt{ecall\_init}).
%\sys initialization is summarized in algorithm \ref{alg:sgxrominit} (\texttt{init\_sgx\_romulus} and \texttt{ecall\_init}).

\begin{algorithm}[!t]
	\caption{--- Initialization algorithms.}
	\label{alg:sgxrominit}
	%\begin{algorithmic}[1]
		%\State $\#\#\#\#\#\#\#\#\#\#\#\#\#\#$\textbf{Out of Enclave}$\#\#\#\#\#\#\#\#\#\#\#\#$
		%\Function{init\_sgx\_romulus}{pm\_file}
		%\State $mapped\_addr \leftarrow mmap (pm\_file)$
		%\State $header\_addr \leftarrow create\_header (mapped\_addr)$
		%\State $ecall\_init(header\_addr)$	
		%\EndFunction
		%\Function{ocall\_unmap}{}
		%\State $munmap(pm\_file)$
		%\EndFunction		
		%\State $\#\#\#\#\#\#\#\#\#\#\#\#\#\#\#$\textbf{In Enclave}$\#\#\#\#\#\#\#\#\#\#\#\#\#\#\#$
		%\Function{ecall\_init}{header\_addr}
		%\State $initialize\_main\_and\_back(header\_addr)$
		%\State $recover()$\cite[p.~5]{romulus}
		%\EndFunction
\begin{lstlisting}
## (*@\color{red}\it Untrusted (outside of enclave)@*) ##
function init_sgx_romulus(pm_file)
	mapped_addr = mmap(pm_file) 
	header_addr = create_header(mapped_addr) 
	ecall_init(header_addr)
end 
function ocall_unmap
	munmap(pm_file)
end
## (*@\color{red}\it Trusted (inside enclave)@*) ##
function ecall_init(header_addr)
	initialize_main_and_back(header_addr)
	recover() (*@\hfill@*) // defined in(*@\cite{romulus}@*)
end function		
\end{lstlisting}
		%\Function{recover}{}
		%\State $pm\_state \leftarrow load(state)$
		%\If{$state == IDLE$} 
		%\State \Return
		%\ElsIf{$state == COPY$}
		%\State $copyMainToBack()$
		%\ElsIf{$state == MUTATING$}
		%\State $copyBackToMain()$
		%\EndIf
		%\State $pfence()$
		%\State $pm\_state \leftarrow IDLE$
		%\EndFunction
	%\end{algorithmic}
\end{algorithm}

\textbf{Initial dataset loading to \pmem.}
\label{subsec:dataload}
One key aspect of \sys is the ability to use training data in \pmem. In \sys, we load training data into PM once, after which the data stays in (byte addressable) \pmem. At resumption following a power failure or system crash, training data in \pmem is instantly accessible to the training algorithm, unlike in disk or SSD-based systems where data needs to be re-read from slow secondary storage into DRAM.

Initially, the training dataset is stored encrypted as files on secondary storage. 
Darknet training algorithms process input data as multidimensional arrays or matrices. 
The goal of this step is to load training data into such a data matrix in \pmem. 
The \emph{sgx-darknet-helper} reads initial training data and labels from secondary storage into DRAM as a volatile matrix variable. 
The address of this matrix is sent to \sgxdnet via an \texttt{ecall}. 
The \emph{pm-data-module} creates a corresponding persistent matrix on \pmem using the \texttt{lib-sgx-romulus} API. 
%For easier integration, the persistent matrix has exactly the same structure as that used by Darknet, except that it is persistent. 
We annotate all persistent types (\eg, matrix rows, matrix values, model layer attributes, \etc.) with the \texttt{persist<>} class from \texttt{lib-sgx-romulus}.
This wrapper class ensures every \texttt{store} operation on the associated persistent data is followed by a \emph{persistent write back} (PWB) to flush the cache line to \pmem. An appropriate fence instruction is used when ordering is required (e.g at the end of a transaction).\footnote{Romulus supports 3 PWB + fence combinations: \texttt{clwb+sfence}, \texttt{clflushopt+sfence} (used in \sys) and \texttt{clflush+nop}.} %\vs{which one do you use in the following?}
Once the persistent matrix is created, the training data is simply \texttt{memcpy}-ied from DRAM into \pmem within a transaction from within the enclave.
%We note that, while this process happens in the enclave runtime, it is relatively inexpensive in terms of enclave memory usage because it transfers data from DRAM to \pmem. 
The persistent data can then be accessed directly via its address.\\ % specifying its address to a routine which requires the former.

\begin{algorithm}[!t]
\caption{--- Training a \ml model in \sys}
\label{alg:training}
%\begin{algorithmic}[1]
%\Function{train\_model}{config}
%\State $enclave\_model \gets create\_enclave\_model(config)$
%\If{$not\_exists(pm\_data)$}
%\State $load\_data\_in\_pm()$
%\EndIf
%\State $iter \gets 0$
%\If{$exists(pm\_model)$}
%\State $mirror\_in(enclave\_model)$
%\State $iter \gets pm\_model.iter$
%\Else
%\State $pm\_model\gets alloc\_mirror\_model(enclave\_model)$
%\EndIf
%
%\Comment{we train for MAX\_ITER iterations}
%\While {$iter < MAX\_ITER$}
%\State $data\_batch \gets decrypt\_pm\_data(batch\_size)$\label{line:databatch}
%\State $train(enclave\_model,data\_batch)$
%\State $mirror\_out(enclave\_model,iter)$
%
%\EndWhile
%\State $free(enclave\_model)$
%\State $ocall\_unmap$
%
%\EndFunction
	
%	\end{algorithmic}
	
\begin{lstlisting}
function train_model(config)
	enclave_model = create_enclave_model(config) 
	if not_exists(pm_data) then
		ocall_load_data_in_pm() 
	end if
	iter = 0
	if exists(pm_model) then
		mirror_in(enclave_model)
		iter = pm_model.iter 
	else
		pm_model = alloc_mirror_model(enclave_model)
	end if
	while iter < MAX_ITER do (*@\hfill@*) // train for max_iter iterations 
		data_batch = decrypt_pm_data(batch_size) 
		train(enclave_model,data_batch) 
		mirror_out(enclave_model,iter)
	end while
	free(enclave_model) 
	ocall_unmap
end function
\end{lstlisting}
\end{algorithm}

\textbf{Model training and mirroring.}
The \sys architecture fits well for training neural network models\cite{mohassel2017secureml}, creating a secure model in enclave memory.
%To do so, \sys creates a secure neural network model in enclave memory.
The architecture of the model and its hyper-parameters (\eg, layer types, batch size, learning rate, \etc) are defined in a config file which is parsed into a \texttt{config} data structure by \emph{sgx-darknet-helper} in the untrusted runtime. Its address is sent to the enclave via an \texttt{ecall} where it is used to build the enclave model.
%To facilitate data sharing across trusted and untrusted worlds, we allocate a \texttt{C struct} in the untrusted runtime, which stores the addresses of the \texttt{config} data structure and data matrix. %which has pointer attributes to useful variables such as the config data structure, data matrix, etc. 
%The address of this \texttt{C struct} is shared once with the enclave prior to training via an \texttt{ecall}.
%The enclave then reads the \texttt{config} data structure to build the enclave model. 
If the training dataset has not been loaded in \pmem, an \texttt{ocall} is performed to load data once from secondary storage into a volatile data matrix variable accessible by the enclave runtime; this could be done in batches if the training dataset is very large. The training data is then loaded into \pmem (see \S\ref{subsec:dataload}).%The address of this matrix is copied to the above mentioned structure in the \texttt{ocall} before the \texttt{ocall} returns. Once inside the enclave, data is loaded as described in \S\ref{subsec:dataload}.

If a persistent mirror model exists on \pmem, we \emph{mirror-in} (read from \pmem and decrypt in enclave) its parameters into the enclave model, otherwise we allocate one in \pmem.
Neural network models in general consist of multiple processing layers with learnable parameters (\ie, weights and biases).
Darknet tracks these layer addresses in an array.
In \sys, we represent a neural network model on \pmem as a linked list of persistent layer structures, so as to simplify future modifications to the model's structure (\eg, add or remove layers). % in the future. 
The model's layers contain persistent attributes,\eg weight vector, bias vector, \etc.
These attributes are annotated with the \texttt{persist<>} class~\cite{romulus}, which ensures PWBs are done for all stores to the corresponding persistent data.

Algorithms~\ref{alg:training} and~\ref{alg:mirroring} summarize respectively model training and mirroring in \sys. 
During model training, batches of training data are decrypted from \pmem (Algorithm~\ref{alg:training}, line 15) into enclave memory and used to train the enclave model for one training iteration. 
After each training iteration we do a \emph{mirror-out} (encrypt in enclave and write to \pmem) of the enclave model parameters to its persistent mirror copy on \pmem.
In the event of a crash during training, upon resumption the model and training data are already in \pmem and can be quickly \texttt{memcpy}-ied from \pmem into secure enclave memory. This obviates the need for much more slower reads from storage devices like SSDs and HDDs.

%transer this to main.tex after lock released
%\begin{lstlisting}[language=C++, caption={Persistent model in \sys}]
%class NVModel{ 

%struct Layer {
%	persist<int> id;       
%	persist<Layer*>next; 
%	persist<uint8_t*>nvweights;
%	persist<uint8_t*>nvbiases;
%	persist<uint8_t*>nvscales;
%	persist<uint8_*>rollxmean;
%	persist<uint8_t*>rollxvar;
%};
%persist<int> num_layers{0};
%persist<size_t> epoch{0};
%persist<float> aloss{0.0};      
%persist<Layer *> head{nullptr}; 

%public:

%NVModel();
%void allocator(network*net);
%void mirror_in(network*net,float*avg_loss);
%void mirror_out(network*net,float*avg_loss);
%};

%\end{lstlisting}

\begin{algorithm}[!t]
\caption{Mirroring algorithms.}
\label{alg:mirroring}
%\begin{algorithmic}[1]
%\Function{alloc\_mirror\_model}{enclave\_model}
%\State $BEGIN\_TRANSACTION$\cite[p.~5]{romulus}
%\State $head\_pm\_L \gets PMalloc(size) $
%\State $head\_pm\_L.W \gets PMalloc(size) $
%\State \Comment{L represents a neural network layer}
%\State \Comment{W represents the layer's parameters}
%\State $head\_pm\_L.next \gets nullptr $
%\State $cur\_pm\_L \gets head\_pm\_L$
%\State $n \gets enclave\_model.numL$
%\For{$i \gets 2$ \textbf{to} $n$}
%\State $cur\_pm\_layer.next \gets PMalloc(size)$ 
%%\Comment{size=size of corresponding enclave layer}
%\State $cur\_pm\_L \gets cur\_pm\_L.next$
%\State $cur\_pm\_L.W \gets PMalloc(size)$ 
%\State $cur\_pm\_L.next \gets nullptr$	
%
%\EndFor
%\State $END\_TRANSACTION$\cite[p.~5]{romulus}
%\EndFunction
%\Function{mirror\_out}{enclave\_model, iter}
%\State $BEGIN\_TRANSACTION$
%\State $n \gets enclave\_model.numL$
%\State $pm\_model.iter \gets iter$
%\State $temp \gets head\_pm\_L$
%\For{$i \gets 1$ \textbf{to} $n$} 
%\State $temp.W \gets $$ encrypt(enclave\_model.L(i).W)$
%\State $temp \gets temp.next$
%\EndFor
%\State $END\_TRANSACTION$
%\EndFunction
%
%\Function{mirror\_in}{enclave\_model}
%\State $BEGIN\_TRANSACTION$
%\State $n \gets pm\_model.numL$
%\State $temp \gets head\_pm\_L$
%\For{$i \gets 1$ \textbf{to} $n$} 
%\State $enclave\_model.L(i).W \gets $$ decrypt(temp.W)$
%\State $temp \gets temp.next$
%\EndFor
%\State $END\_TRANSACTION$
%\EndFunction
%\end{algorithmic}

\begin{lstlisting}
function alloc_mirror_model(enclave_model) 
	BEGIN_TRANSACTION (*@\hfill@*) // defined in(*@\cite{romulus}@*)
	head_pm_L = PMalloc(size) (*@\hfill@*) // L: neural network layer 
	head_pm_L.W = PMalloc(size) (*@\hfill@*) // W: layer's parameters
	head_pm_L.next = nullptr 
	cur_pm_L = head_pm_L
	n = enclave_model.numL 
	for i = 2 to n do
		cur_pm_L.next = PMalloc(size) 
		cur_pm_L = cur_pm_L.next 
		cur_pm_L.W = PMalloc(size) 
		cur_pm_L.next = nullptr
	end for
	END_TRANSACTION (*@\hfill@*) // defined in(*@\cite{romulus}@*)
end function		
function mirror_out(enclave_model,iter) 
	BEGIN_TRANSACTION
	n = enclave_model.numL 
	pm_model.iter = iter
	temp = head_pm_L 
	for i = 1 to n do
		temp.W  = encrypt(enclave_model.L(i).W)
		temp = temp.next 
	end for
	END_TRANSACTION 
end function
function mirror_in(enclave_model) 			
	n = pm_model.numL
	temp = head_pm_L
	for i = 1 to n do
		enclave_model.L(i).W = decrypt(temp.W)
		temp = temp.next			
	end for		
end function
\end{lstlisting}
\end{algorithm}

%!TEX root = main.tex
\section{Evaluation}
\label{sec:evaluation}

Our experimental evaluation of the \sys prototype answers the following questions:
\begin{itemize}[leftmargin=*]
	%\item Why we designed \sgxrom instead of using systems such as SCONE which allow to run unmodified applications (thus \pmem libraries) inside SGX enclaves?
	\item How \sgxrom compares against unmodified Romulus in a SCONE container ?
	\item How does \sys improve checkpoint/restore performance when compared to secondary storage (e.g SSD)?
	\item How scalable is \sys when varying model sizes?
	\item What are the main bottlenecks in the \sys design?
	\item What is the overhead of batched-data decryptions?
	\item Is the mirroring mechanism robust against crashes?	
	\item Are there processing and storage bottlenecks?	
\end{itemize}

\textbf{Experimental setup.}
At the time of this writing (October 2020), servers that support both SGX and \pmem are not available. 
Hence, we perform our experiments on two different servers, \ie, \emph{sgx-emlPM} and \emph{emlSGX-PM}.
The \emph{sgx-emlPM} node supports SGX but has no physical \pmem, hence we resort to emulating the latter with Ramdisk. %url{https://pmem.io/}}.
This machine is equipped with a quad-core Intel Xeon E3-1270 CPU clocked at 3.80\,GHz, and 64\,GB of DRAM.
The CPU ships with 32\,KB L1i and L1d caches, 256\,KB L2 cache and 8\,MB L3 cache.
Concerning \emph{emlSGX-PM}, it is equipped with 4$\times$ Intel OptaneDC DIMMs of 128\,GB each. %\pmem.
However its processors lack native support for SGX. 
Hence, we resort to SGX in simulation mode~\cite{sgxdevref}.
The \emph{emlSGX-PM} node is a dual-socket 40-core Intel Xeon Gold 5215 clocked at 2.50\,GHz and 376\,GB of DRAM.
Each processor has 32\,KB L1i and L1d caches, 1\,MB L2 cache and a shared 13.75\,MB L3 cache. 
Both servers run Ubuntu 18.04.1 LTS 64\,bit and Linux kernel 4.15.0-54. 
We run the Intel SGX platform software, SDK and driver version v2.8.
All our enclaves have max heap sizes of 8\,GB and stack sizes of 8\,MB. The EPC size is 128\,MB (93.5\,MB usable). 
Unless stated otherwise, we use \texttt{CLFLUSHOPT} and \texttt{SFENCE} for persistent write backs and ordering. % respectively in our experiments.

By using both servers, we highlight the performance implications of both real SGX and real PM.
All experimental comparisons are executed separately for each server, as they have completely different characteristics. We indicate where necessary on which node an experiment is carried out.
All SCONE containers are based on Alpine Linux~\cite{kunkel2019tensorscone}.

All models used in our evaluations are convolutional neural networks (CNNs).
The convolutional layers use \emph{leaky rectified linear unit} (LReLU)~\cite{grover2018privado} as activation, and all output layers are \emph{softmax}~\cite{mohassel2017secureml} layers.
The model optimization algorithm used is stochastic gradient descent (SGD), and the learning rate used is 0.1.
Except stated otherwise, all training iterations use a batch size of 128. 
Concerning the dataset, we use MNIST~\cite{mnist}, a very popular dataset in the deep learning community. It consists of 70'000 grayscale images of handwritten digits (60'000 training samples and 10'000 test samples).

\begin{figure}[t!]
	\centering
	%\hspace*{-0.75cm}
	\includegraphics[scale=0.68,trim={0 0 0 0}]{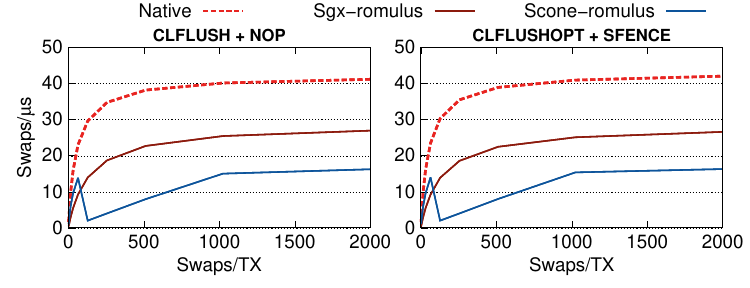}
	\caption{SPS benchmark on the \emph{sgx-emlPM} with varying transaction sizes for two PWB and fence combinations: \texttt{CLFLUSH+ NOP} (left) and \texttt{CLFLUSHOPT + SFENCE} (right).}
	\label{fig:sps}		
\end{figure}

\textbf{Why SGX-Romulus makes sense.}
We begin by comparing \sgxrom with the unmodified Romulus library running in a SCONE container, with the goal of understanding how a manually ported library using the Intel SGX SDK and the unmodified version in SCONE stand against each other.
%Our goal is to understand the performance differences between our manually ported library in the Intel SGX SDK and the unmodified version running in a SCONE container. 

We measure how many swaps per second (SPS) they achieve, a metric commonly used~\cite{romulus} to compare PM libraries. 
SPS stores an array of integers in \pmem and evaluates the overhead of randomly swapping array values within a transaction, for different \emph{persistence fences} and transaction sizes. 
%We use it to additionally evaluate the overhead of using fences from within SGX enclaves.
%In our case, it gives us an additional evaluation of the overhead of fences within SGX enclaves. 
This experiment uses the \emph{sgx-emlPM} node, as real SGX is the main factor that dictates the performance differences.  %between the systems compared. 
For each transaction size we run SPS for 20\,s.
Figure~\ref{fig:sps} shows the throughput of swap operations on a 10\,MB persistent array with different transaction sizes for different systems with a single threaded application. 
We include results for two choices of PWB implemented by Romulus and \sgxrom: \texttt{clflush + NOP} and \texttt{clflushopt + sfence}. Our servers do not have support for \texttt{clwb}.

We observe that in both cases, the persistence fences take approximately 1.6$\times$ to 3.7$\times$ longer to complete in \sgxrom when compared to native (no SGX) systems for transaction sizes between 2 and 2048 swaps operations per transaction. 
When compared to Romulus in SCONE, transactions for both fence implementations in \sgxrom are approximately 1.5$\times$ to 2.5$\times$ slower for transaction sizes between 2 to 64 swap operations per transaction. 
Beyond 64 swap operations per transaction, there is a pronounced drop in throughput for Romulus in SCONE, and \sgxrom transactions are 1.6$\times$ to 6.9$\times$ faster. 
We justify this behaviour as a result of limited space available for Romulus' volatile redo log in the SCONE container. 
These results suggest \sgxrom is a preferable choice for our \ml system, where multiple operations are carried out on persistent models within transactions of relatively larger sizes.

%As a final remark, a naive implementation of \sys using unmodified libraries in SCONE containers or a libOS (such as Graphene) leads to a very high TCB size. 
%By manually porting \sgxdnet and \sgxrom via separation into trusted and untrusted components, we decrease the total number of lines of code (LOC) in the enclave.
%\sys thus achieves a TCB reduction of $\sim$44\% in terms of LOC.

\begin{figure}[t!]
	\centering
	\includegraphics[scale=0.68,trim={0 0 0 0}]{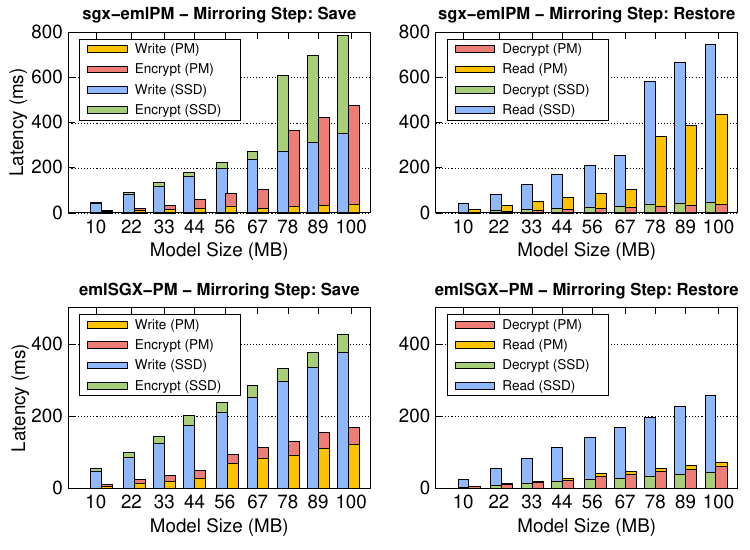}
	\caption{PM mirroring vs. checkpointing on SSD for sgx-emlPM (top) and emlSGX-PM (bottom).}%\vs{change colors.}%\py{keys have a problem: top keys are valid only for saves and bottom keys only for restores.}
	\label{fig:realsgx-pm}	
\end{figure}

\textbf{PM mirroring vs. SSD-based checkpointing.} Next, we compare the mirroring mechanism in \sys to traditional checkpointing on a SSD using \sgxdnet.
For SSD checkpointing, we use \texttt{ocall}s to \texttt{fread} and \texttt{fwrite} libC routines to read/write from/to SSD.
After each call to \texttt{fwrite}, we flush the libC buffers and issue an \texttt{fsync}, to ensure data is actually written to secondary storage.
We vary model sizes by increasing the total number of convolutional layers.
We measure the times to save/\texttt{mirror-out} (encrypt in the enclave and write to PM) and restore/\texttt{mirror-in} (read from PM into enclave and decrypt), and compare these to SSD-based checkpoint saves (encrypt and write to SSD) and SSD-based checkpoint restores (read from SSD into enclave and decrypt), which are the state-of-the-art methods for fault tolerance.
All data points are an average of 5 runs.

Figure~\ref{fig:realsgx-pm} represents the results obtained on our two servers.
As a general observation, in \sys, in-enclave data encryption contributes more to the save-latency (\ie, mirror-out) when compared to writes to \pmem.
For restores in \sys, reads from \pmem into enclave memory contribute more to the overall latency.
When compared to traditional saves and restores on SSD, our mirroring mechanism gives less overhead.
%We present a more in-depth analysis of these results in the rest of this section. 
%\vs{these results are not described: put 2/3 sentences}\py{I don't understand what you mean..}

\cref{tab:steps} shows a performance breakdown of each mirroring steps for saves and restores in \sys, while Table~\ref{tab:pmdisk} shows the average performance improvements of our mirroring mechanism when compared to SSD-based checkpointing.
To reduce the effect of outliers, we evaluate results beneath and beyond the EPC limit separately.
The usable EPC size is 93.5\,MB, reached for model size 78\,MB, due to the presence of other data structures in enclave memory (\eg, temporary buffers used for encryption) as well as enclave code.
We observe (Table~\ref{tab:steps}) that for saves in a real SGX environment, encryption contributes more (66.4\%) to the overall mirroring latency on average for model sizes beneath 78\,MB. 
This jumps to 92.3\% once the EPC limit is crossed.
This overhead is due to expensive page swapping between the EPC and regular DRAM by the SGX kernel driver.
For restores, reads contribute on average 75\% and 91.2\% for values beneath and beyond the EPC limit respectively.
Similarly, we have a high overhead beyond the EPC limit due to the SGX driver's page swaps. 
Our results show in-enclave decryption is relatively cheaper.

\begin{table}[b!]
	\setlength{\tabcolsep}{2pt}
	\scriptsize
	\begin{subtable}{.48\linewidth}
	\centering
	\caption{Breakdown of mirroring steps (\%)}
	\label{tab:steps}
 	% \begin{tabular}{l@{\hskip 0pt}|c@{\hskip 0pt}|c@{\hskip 0pt}}\small
	\begin{tabular}{l|c|c}
	\textbf{Save} & \textbf{SGX-emlPM} & \textbf{emlSGX-PM}\\
	\hline
	\multirow{2}{*}{Encrypt} & 66.4\% & \multirow{2}{*}{30.3\%}\\
	\cline{2-2}
	 &\cellcolor{lightgray} 92.3\%&\\
	\hline{}	
	\multirow{2}{*}{Write} & 33.6\% & \multirow{2}{*}{69.7\%}\\
	\cline{2-2}
	&\cellcolor{lightgray}7.7\%&\\
	\hline 
	\hline
	\textbf{Restore} & \textbf{A} & \textbf{B}\\
	\hline
	\multirow{2}{*}{Read} & 75\% & \multirow{2}{*}{17.8\%}\\
	\cline{2-2}
	&\cellcolor{lightgray}91.2\%&\\
	\hline 
	\multirow{2}{*}{Decrypt} & 25\% & \multirow{2}{*}{82.2\%}\\
	\cline{2-2}
	&\cellcolor{lightgray}8.8\%&\\
	\hline 		
	\end{tabular}%
	\end{subtable}%
\hspace{\fill}
	\begin{subtable}{.48\linewidth}
	\centering
	\caption{\sys speed-ups}
	\label{tab:pmdisk}

	% \begin{tabular}{l@{\hskip 0pt}|c@{\hskip 0pt}|c@{\hskip 0pt}}\small
	\begin{tabular}{l|c|c}
		\textbf{Save} & \textbf{SGX-emlPM} & \textbf{emlSGX-PM}\\
		\hline
		\multirow{2}{*}{Write} & 7.9$\times$ & \multirow{2}{*}{4.5$\times$}\\
		\cline{2-2}
		&\cellcolor{lightgray}9.6$\times$&\\
		\hline 
		\multirow{2}{*}{Total} & 3.5$\times$ & \multirow{2}{*}{3.2$\times$}\\
		\cline{2-2}
		&\cellcolor{lightgray}1.7$\times$&\\
		\hline 
		\hline
		\textbf{Restore} & \textbf{A} & \textbf{B}\\
		\hline
		\multirow{2}{*}{Read} & 3$\times$ & \multirow{2}{*}{16.8$\times$}\\
		\cline{2-2}
		&\cellcolor{lightgray}1.8$\times$&\\
		\hline 
		\multirow{2}{*}{Total} & 2.5$\times$ & \multirow{2}{*}{3.7$\times$}\\
		\cline{2-2}
		&\cellcolor{lightgray}1.7$\times$&\\
		\hline 		
	\end{tabular}%		
	\\

	\end{subtable}
		\centering
	\caption{Shaded cells: values beyond the EPC size.}
	%\textbf{A = server SGX-emlPM} and \textbf{B = server emlSGX-PM}.
\end{table}

For the \emph{emlSGX-PM} server, without real SGX hardware (hence no expensive page swaps), the main bottleneck is real \pmem.
We observe (\cref{tab:pmdisk}) that for server \emph{sgx-emlPM}, writes to \pmem are on average 7.9$\times$ and 9.6$\times$ faster when compared to writes to SSD for enclave sizes beneath and beyond the EPC limit respectively. 
SSD writes are generally more expensive due the expensive \texttt{ocall}s and serialization operations to secondary storage. 
Saves are overall 3.5$\times$ and 1.7$\times$ faster for enclave sizes beneath and beyond the EPC limit respectively. 
Similarly, for restores, reads from \pmem into enclave memory are on average 3$\times$ and 1.8$\times$ faster for enclave sizes beneath and beyond the EPC limit respectively, when compared to the SSD-based counterpart.
Restores are overall 2.5$\times$ and 1.7$\times$ faster for enclave sizes beneath and beyond the EPC limit.
A similar breakdown is done for the \emph{emlSGX-PM} node. 

\textbf{Training larger models.} Our results suggest \sys is best suited for models with sizes beneath the EPC limit. Models larger than the EPC limit can be trained with \sys but this leads to a significant drop in training performance due to the extensive page swaps by the SGX kernel driver. Figure \ref{fig:realsgx-pm} shows our mirroring mechanism still peforms better than SSD-based checkpointing for model sizes beyond the EPC limit. A possible strategy to overcome the EPC limitation could be to distribute the training job over multiple secure CPUs. We will explore this idea in the future. Also, a recent processor release by Intel expands the EPC to 256\,MB~\cite{itpeer}. This paves the way for applications that leverage \sys to train much larger models more efficiently.

\textbf{Mirroring frequency.} By default \sys does mirroring after every iteration. The mirroring frequency can be easily increased or decreased. All things being equal, a training environment with a small or high frequency of failures will require respectively, small or high mirroring frequencies to achieve good fault tolerance guarantees.

\begin{figure}[t!]
	\centering
	%\hspace*{-0.75cm}
	\includegraphics[scale=0.68,trim={0 0 0 0}]{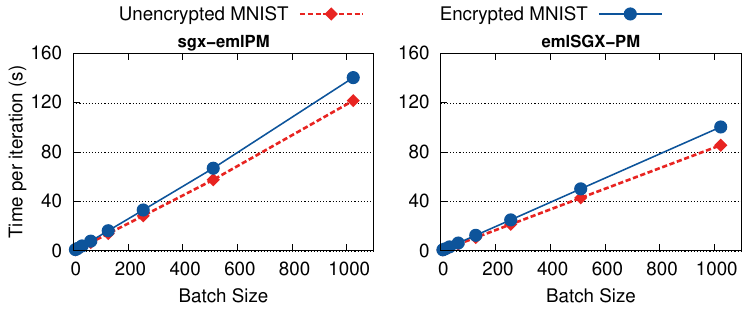}
	\caption{Variation of iteration times with different batch sizes for encrypted and unencrypted MNIST data.}	
	\label{fig:databatch}
\end{figure}

\textbf{Overhead of data batch decryptions.}\label{ssec:batchres}
For efficiency reasons, \ml algorithms (\eg, SGD) manipulate training data in batches for each training iteration. 
In this experiment we study the performance impact on total iteration time of batch decryptions of training data into enclave memory. 
We proceed by comparing the iteration times with different batch sizes for a model being trained via the \sys mechanism, to a model trained with batches of unencrypted data on \pmem.
We recall that in \sys, batches of encrypted training data are read from \pmem and decrypted in enclave memory for each iteration.
All models have 5 LReLU-convolutional layers.

Figure~\ref{fig:databatch} shows the results obtained on both systems. 
We observe that iterations with batch decryption of data into enclave memory are 1.2$\times$ slower on average for both systems. 
We consider this a relatively small price to pay for data confidentiality during training. 

\begin{figure}[t!]
	\centering
	%\hspace*{-0.75cm}
	\includegraphics[scale=0.68,trim={0 0 0 0}]{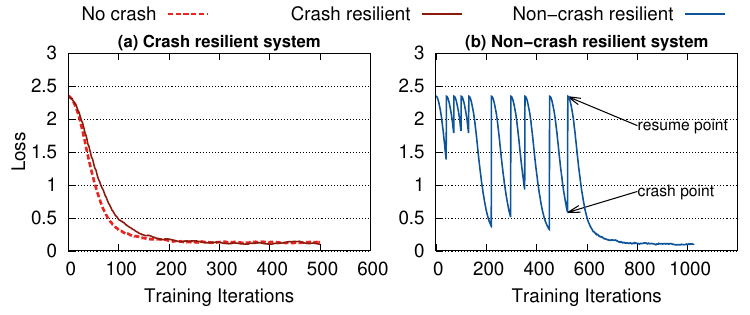}
	\caption{Crash/resumes are done by randomly killing and restarting the training process every 10 to 15 minutes during model training.}	
	\label{fig:crash}
\end{figure}

\textbf{Crash resilience.}\label{ssec:crashres}
The main purpose of our experiments here is to demonstrate that \sys's mirroring mechanism is \emph{crash resilient} (or failure transparent), as well as demonstrate the performance impact on the training process of a \emph{non-crash-resilient} system.
We define a crash-resilient system as one capable of recovering its state (\ie, learned parameters) prior to a system crash.%Otherwise we consider it non-crash-resilient.
The experiments consider models with 5 LReLU-convolutional layers, trained with the MNIST dataset for 500 iterations.
We study the variation of the loss while doing random crashes during model training.
 
Figure~\ref{fig:crash} presents the results obtained on the \emph{emlSGX-PM} server, but similar results are obtained on \emph{sgx-emlPM}.
We proceed by training a model using \sys with 9 random crashes (and resumptions) during the training process.
We compare the loss curve obtained here to one obtained without any crashes (baseline).
Figure~\ref{fig:crash}(a) shows that despite the crashes, the loss curve follows closely (no breaks at crash and resume points) the one obtained without crashes.
This indicates the model parameters are saved and restored correctly using the mirroring mechanism in \sys.
In comparison, Figure~\ref{fig:crash}(b) shows the loss curve obtained when the system cannot recover its learned parameters following random crashes.
For this experiment we run our system while disabling model's weights saving via our mirroring mechanism.
At every resumption point, the model begins the learning process with initial randomized weights, and thus still requires 500 iterations to be fully trained, hence increasing the total iterations (from when training first began) required to train the model to over 1000 in this experiment.
This shows the benefit of crash-resilience in an \ml system. In the next section, we use a more realistic crash/resume pattern (spot instance trace) to show crash resilience in \sys.
%\vs{ I would move this section on crash resilience after the batch data decryption}

\begin{figure}[t!]
	\centering
	\includegraphics[scale=0.68,trim={0 0 0 0}]{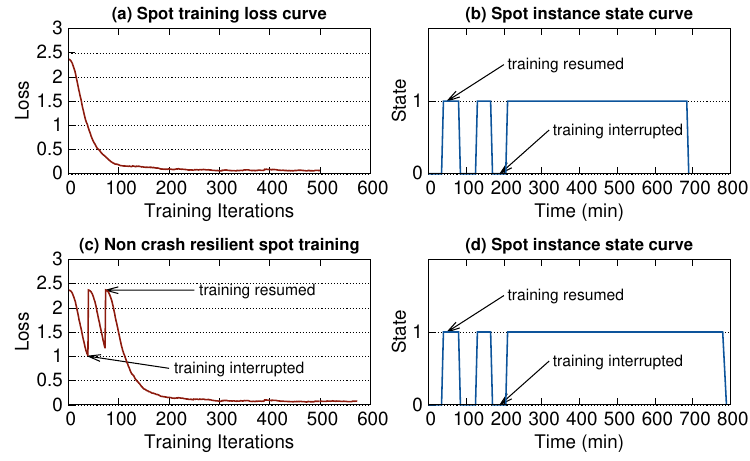}
	\caption{Model training with AWS EC2 spot instance traces.}%\vs{change colors.}%\py{keys have a problem: top keys are valid only for saves and bottom keys only for restores.}
	\label{fig:spot}
\end{figure}

\textbf{\sys on AWS EC2 Spot instances.}
A practical use case for \sys framework would be model training on spot instances, such as those offered by Amazon EC2 and Microsoft Azure. Spot instances are liable to many interruptions during their lifetimes, and model training in such a scenario requires efficient fault tolerance guarantees (such as those provided by \sys) to reduce cost and increase efficiency of the training process.
We use Amazon EC2 spot instance traces from~\cite{wang2018empirical} to simulate a realistic model training scenario with \sys on a spot instance.
The spot traces contain market prices of spot instances at different timestamps (5 minutes intervals).
To simulate spot model training, we set a \emph{maximum bid price} in our simulator script, and our simulation algorithm periodically (every 5 minutes) compares the \emph{market price} at each timestamp in the spot trace to our bid price.
If $max\_bid > market\_price$, our training process is launched (or continues if it was already running).
Otherwise, the training process is killed.
We train a model with 12 LReLU-convolutional layers for 500 iterations on server \emph{emlSGX-PM}.

Figure~\ref{fig:spot}(a) shows the loss curve obtained after 500 iterations.
As explained in the previous section, this shows \sys is crash resilient as training resumes where it left off prior to the training process being stopped.
Figure~\ref{fig:spot}(b) shows a ``state curve'' of the training process (or spot instance) throughout the training process.
The process state is 1 when it is running and 0 otherwise.
We observe only 2 interruptions of the training process with our simulation parameters (\ie, maximum bid price of 0.0955).
The chosen maximum bid price and spot market price variations will dictate the total number of interruptions of the spot instance, and hence the total training time (interruption times included).
The spot traces used and our simulation scripts are available in the \sys repository. 

Figure~\ref{fig:spot}(c) shows us the loss curve obtained when there is no crash resilience (\ie, the model's state is not saved).
With the given simulation parameters (\ie, maximum bid price of 0.0955), there are two interruptions during the training process.
As explained in the previous section, it needs to resume training afresh, and hence the combined number of iterations (and total time) from when training first began is increased when compared to its crash resilient counterpart.
This further justifies the need for fault tolerance guarantees in such \ml scenarios.

%\subsection{Some Limitations of our framework}
%\paragraph{Lack of GPU} The absence of SGX on GPUs means the latter cannot be leveraged for accelerating complex \ml training operations. 
%
%\paragraph{SGX threading model} SGX enclaves cannot spawn threads; threads can only be created in the untrusted runtime. This poses some difficulties for multithreaded training in enclaves. \ml libraries can however be carefully designed to leverage the benefits of parallelism inside enclaves. This could be an interesting extension for future work.  

\textbf{CPU and memory overhead.}
Our mirroring mechanism uses 140 bytes of \pmem for encryption metadata per layer. 
The MAC is 16\,B, the IV is 12\,B, giving 28\,B per encrypted parameter buffer. 
Each layer contains 5 parameter matrices, hence $28\times5 = 140$\,B per layer. 
With a model of $N$ layers, we account for $N\times140$ extra bytes on PM for encryption metadata, small compared to the size of actual models (order of few MBs).
The training algorithm is a fairly intensive single-threaded application and it uses 98-100\% of the CPU during execution.

\textbf{Secure inference.}
%While \sys is mainly adapted for secure training of \ml models, it also supports secure inference.
\sys can also be used for secure inference. We trained a CNN model with 12 LReLU convolutional layers on the MNIST training dataset, and used the trained model to classify 10'000 grayscale images of handwritten digits in the range $[0-9]$. 
The model (available in the \sys repository) achieved an accuracy of 98.52\% with the given hyper-parameters. 
%The complete model architecture used is available in the \sys repository.

\textbf{GPU and TPU support.}
Hardware accelerators like Graphics Processing Units (GPUs) and Tensor Processing Units (TPUs) are increasingly used in \ml applications. However the former do not integrate TEE capabilities. Recent works like HIX~\cite{hix}, Graviton~\cite{graviton}, and Slalom~\cite{slalom} propose techniques to securely offload expensive \ml computations to GPUs. Using Darknet's CUDA extensions, \sys can leverage such techniques to improve training performance. The trained model weights can be securely copied between the secure CPU and the GPU (or TPU) and our mirroring mechanism applied without much changes. We are exploring possible improvements of \sys in this direction.

%!TEX root = main.tex
%\vspace{-2pt}
\section{Related work}
\label{sec:related}
\vspace{-2pt}
\smallskip\noindent\textbf{TEE-based schemes.}
There exists several solutions leveraging trusted hardware (\ie, Intel SGX) for secure \ml.
Slalom~\cite{slalom} is a framework for secure DNN inference in TEEs.
It outsources costly neural network operations to a faster, but untrusted GPU during inference.
Occlumency~\cite{Lee2019} leverages Intel SGX to preserve confidentiality and integrity of user-data during deep learning inference in untrusted cloud infrastructure.
Privado\cite{grover2018privado} implements a secure inference-as-a-service, by
eliminating input-dependent access patterns from \ml code, hence reducing data leakage risks in the enclave.
Chiron~\cite{chiron} leverages Intel SGX for secure ML-as-a-service which prevents disclosure of both data and code. 

These systems leverage TEEs for model inference, but without any support for failure recovery. 
\sys provides a full framework that supports both in-enclave model training and inference with efficient fault tolerance guarantees on \pmem. 

SecureTF~\cite{kunkel2019tensorscone} integrates TensorFlow \ml library for model training and inference in secure SCONE containers. This requires the full TensorFlow library (over 2.5 million LOC~\cite{openhub}) to run inside SGX enclaves, which by design increases the TCB. On the other hand, the trusted portion of \sys comprises 15'900 LOC. The reduction in TCB in \sys when compared to SecureTF is quite obvious; this is better from a security perspective.

% showing feasible in-enclave model training.
%To the best of our knowledge, we are the first to propose mechanisms to leverage \pmem in TEEs for fault tolerance in the context of privacy-preserving \ml.

%\vspace{0.25cm}
\smallskip\noindent\textbf{Homomorphic encryption (HE)-based schemes.}
Without trusted hardware enclaves, many privacy-preserving \ml methods achieve security via HE-based techniques. 
HE schemes compute directly over encrypted data. % without the need to decrypt the latter.
CryptoNets~\cite{gilad2016cryptonets} implements inference over encrypted data for pre-trained neural networks. 
Solutions exist~\cite{hesamifard2017privacy} to train and do inference on neural network models using HE.
%Hesamifard et al.~\cite{hesamifard2017privacy} and Aslett et al.~\cite{aslett2015encrypted} provide solutions to train and do inference on neural network models using HE. 
%Gazelle\cite{gazelle} is a low-latency framework for secure neural network inference using a combination of HE and two-party computation techniques.

While these methods ensure privacy of sensitive training and classification data during model training and inference, they have significant performance overhead (up to 1000$\times$ slower than TEE-based schemes~\cite{hynes2018efficient}). 
%Practical techniques for training complex models over deep neural networks via HE schemes are yet to be developed~\cite{chiron}. 
\sys provides an orthogonal approach to tackle security, combining Intel SGX enclaves to ensure confidentiality and integrity of models and data sets during training and inference at a much lower cost. 
%Furthermore, \sys provides novel mechanisms which leverage \pmem for fault tolerance during model training in TEEs. %% said before already

%\vspace{0.25cm}
\smallskip\noindent\textbf{Fault tolerance in \ml.}
A common technique for fault tolerance in \ml learning frameworks is checkpointing (restoring) of the model's state to (from) secondary storage during training (recovery). 
%Upon resuming after a system failure, the most recent checkpoint is loaded into main memory as starting point for the training phase. 
%These systems also rely on secondary storage as persistent storage for training data throughout the training process.
Several frameworks (\ie, Tensorflow~\cite{abadi2016tensorflow}, Caffe~\cite{caffe},  Darknet~\cite{darknet}, \etc) rely on secondary storage as persistent storage for training data throughout the training process.
Distributing training across several compute nodes improves scalability while increasing fault tolerance. % while increasingly but it is primarily geared towards performance.

The above mentioned techniques have huge performance overhead, due to high access times of secondary storage. 
Following a crash, entire data sets and models must be reloaded into main memory from secondary storage.
\sys's novel mirroring mechanism leverages \pmem for fault tolerance: upon a crash, the model and the associated training data are readily available in memory. 
Our design completely obviates the need for expensive serialization (deserialization) of models to (from) secondary storage, and proposes a more efficient approach for handling large amounts of training data. 

%LMC~\cite{vogt15dsn}.
%ReplicaTEE~\cite{soriente2019replicatee}. 
%Private Neural Network Inference in Intel SGX~\cite{Federschmidt2019}

%!TEX root = main.tex
\section{Conclusion}
\label{sec:conclusion}

\sys is the first secure \ml framework to leverage Intel SGX for secure model training and \pmem for fault tolerance. 
Our novel mirroring mechanism creates encrypted mirror copies of enclave \ml models in \pmem, which are synchronized across training iterations. 
Our design leverages \pmem to store byte-addressable training data, completely circumventing expensive disk I/O operations in the event of a system failure. 
The evaluation of \sys shows that its design substantially reduces the TCB when compared to a system with unmodified libraries, and the mirroring mechanism outperforms disk-based checkpointing systems while ensuring the system's robustness upon system failures. 
Using real-world datasets for image recognition, we show that \sys offers a practical solution to securely train \ml models in TEEs integrated with PM hardware at a reasonable cost.

We will extend this work along the following directions. First, we intend to explore GPUs and TPUs by offloading expensive enclave operations on the former without a loss in confidentiality. The extent to which this can be done while
preserving confidentiality of the model parameters and training or inference data will be the key area for future work.
Second, we wish to explore distributed training using \sys to overcome the SGX EPC limitation. Lastly, we plan to better exploit system parallelism to improve the performance of \sys. This entails redesigning \sgxdnet to efficiently support parallel training with threads spawned in the untrusted runtime. 

%as follows: \emph{(i)} by exploring how to offload to GPUs expensive enclave operations without loss in confidentiality for both model parameters and inference data, and \emph{(ii)} by better exploiting system parallelism to improve the performance of \sys.
%The extent to which this can be done while preserving confidentiality of the model parameters and training or inference data will be the key area for future work.
%Second, by better exploiting system parallelism to improve the performance of \sys.
%By design, Intel SGX prohibits the spawning of threads from within an enclave.
%This entails redesigning \sgxdnet to efficiently support parallel training with threads spawned in the untrusted runtime.

%\vs{but how do you integrate with sgx?}.
%\vs{can you explain a bit more what you mean by \emph{parallelism} ? I'll expand the text}.
%\py{Parallelism = multithreading..} \py{as for GPU integration, expensive enclave operations can be outsourced to a more powerful but untrusted GPU with little loss in confidentiality, just as shown in the Slalom paper.}

\section*{Acknowledgment}
This work received funds from the Swiss National Science Foundation (FNS) under project PersiST (no. 178822).

%\section*{References}

%\begin{thebibliography}{00}
{\footnotesize
\bibliographystyle{IEEEtranS}
\bibliography{compact}
}
%\end{thebibliography}
\vspace{12pt}

\end{document}